\documentclass{ws-ijmpcs}
\usepackage{amssymb}
\usepackage{amsmath}
\usepackage{bm}
\usepackage{upgreek}
\usepackage{graphicx}

\begin{document}

\markboth{Authors' Names}
{Instructions for Typing Manuscripts (Paper's Title)}

%
\catchline{}{}{}{}{}
%

\title{CASIMIR FRICTION FORCE FOR MOVING HARMONIC OSCILLATORS
}

\author{JOHAN S. H{\O}YE}

\address{Department of Physics, Norwegian University of Science and Technology, N-7491 Trondheim, Norway
\\
johan.hoye@ntnu.no}

\author{IVER BREVIK}

\address{Department of Energy and Process Engineering, Norwegian University of Science and Technology, N-7491 Trondheim, Norway \\
iver.h.brevik@ntnu.no}

\maketitle

\begin{history}
\received{Day Month Year}
\revised{Day Month Year}
\end{history}

\begin{abstract}
Casimir friction is analyzed for a pair of dielectric particles in relative motion. We first adopt a microscopic model for harmonically oscillating particles at finite temperature $T$ moving non-relativistically with constant velocity. We use a  statistical-mechanical description where time-dependent correlations are involved. This description is physical and direct, and,  in spite of its simplicity, is able to elucidate the essentials of the problem. This treatment elaborates upon, and extends, an earlier theory of ours back in 1992. The energy change $\Delta E$ turns out to be finite in general, corresponding to a finite friction force. In the limit of zero temperature the formalism yields, however, $\Delta E \rightarrow 0$, this being due to our assumption about constant velocity, meaning slowly varying coupling. For couplings varying more rapidly, there will also be a finite friction force at $T=0$. As second part of our work, we consider the friction problem using time-dependent perturbation theory. The dissipation, basically a second order effect, is obtainable with the use of first order theory, the reason being the absence of cross terms due to uncorrelated phases of eigenstates. The third part of the present  paper is to demonstrate explicitly the equivalence of our results with those recently obtained by Barton (2010); this being not a trivial task since the formal results are seemingly quite different from each other.

\end{abstract}

\ccode{PACS numbers: 05.40.-a, 05.20.-y, 34.20.Gj}

\section{Introduction and background}

    \begin{figure}[htbp]
\makebox[\textwidth][c]{
        \includegraphics[width=5cm]{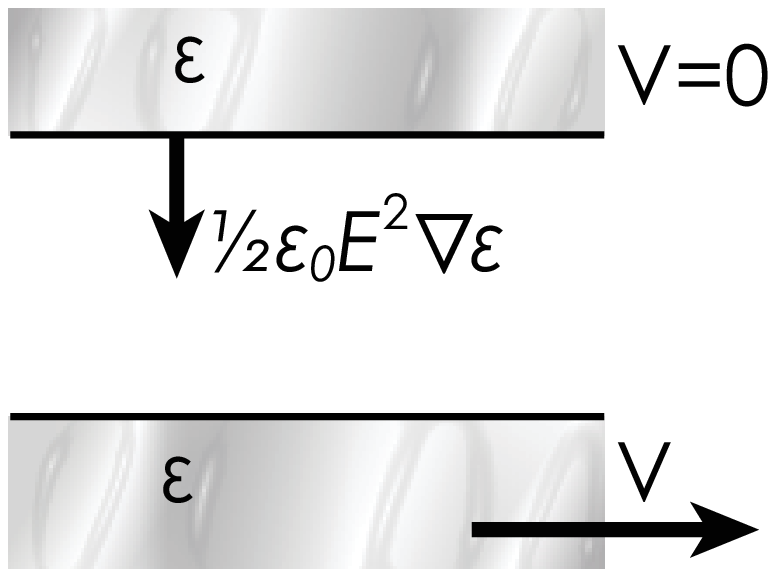}
        }
    \begin{center}
    \end{center}
\end{figure}

Consider the arrangement sketched in Fig.~1, namely two parallel semi-infinite dielectric nonmagnetic plates at micron or semi-micron separation, the upper plate kept at rest, the lower moving with a constant nonrelativistic velocity $V$. For simplicity we assume the media to be of equal composition, their permittivity being real, equal to $\varepsilon$.  In principle, we may allow for dispersion. The electromagnetic force density in the boundary layer of the resting plate is
\begin{equation}
{\bf f}=-\frac{1}{2}\varepsilon_0E^2\nabla \varepsilon. \label{1Q}
\end{equation}
The essential property of the force (\ref{1Q}) in our present context is that it is {\it vertically} directed, i.e. transverse to the relative motion of the plates. And it is precisely this force which is able to explain all the conventional experiments in classical electrodynamics where dielectric media, especially liquids, are involved. Let us for illustration mention three examples:

$\bullet$ The first is the classic radiation pressure experiment of Ashkin and Dziedzic \cite{ashkin73}. Focused light was sent from above towards an air-water surface, and an outward bulge of the surface was observed, of the order of 1 $\mu$m. The light source was a pulsed frequency doubled YAG laser pulse (duration of each pulse  60 ns, peak power 3 kW, beam waist 4.5 $\mu$m). The reason for the smallness elevation is the large surface tension for air-water. Theoretical papers on this experiment can be found in Refs.~\refcite{lai76} and \refcite{brevik79}.

$\bullet$ Our second example is the experimental work of Zhang and Chang \cite{zhang88}, showing clearly how a water droplet becomes deformed when hit by a laser pulse. Typical laser energies were 100 mJ in each pulse. Theoretical treatments for this case are given in Refs.~\refcite{lai89} and \refcite{brevik99}.

$\bullet$ A newer series of experiments is provided by the Bordeaux group \cite{casner04,delville09,wunenburger11}, working with a two-fluid system of surfactant-coated nanodroplets in oil microemulsions near the critical point, implying that the surface tension can be made about $10^6$ times smaller than the usual air-water tension. The surface displacement can accordingly be very large, about 70 $\mu$m.

In our context, the  point to be emphasized is that in all these cases the expression (\ref{1Q}) provides the correct electromagnetic force. And this brings us to the present theme: how can the Casimir {\it friction force} be accounted for, by the use of conventional electrodynamics? Obviously, the transverse force (\ref{1Q}) cannot be of importance for friction. It is clear that we here have to leave the assumption of a real permittivity. Dissipation has to be rooted in energy {\it dissipation}, that means, in a permittivity that is complex valued.

It is worthwhile to show how one can come to the same conclusion via an alternative approach, namely with the use of field theory. Let us start from Schwinger's source theory, giving how the electric field components $E_i$ are related to the polarization source $P_k$ via a generalized susceptibility $\Gamma_{ik}$ \cite{schwinger78}
\begin{equation}
 E_i(x)=\int d^4x \Gamma_{ik}(x,x')P_k(x'). \label{2Q}
 \end{equation}
Stationarity means that we can take the difference $\tau=t-t'$ as time variable.  Causality means that $t' \leq t$. We take the Fourier transform
to obtain the spectral representation $ \Gamma_{ik}({\bf r,r'}, \omega).$
The Kubo formula relates $\Gamma$ to the commutator:
\begin{equation}
\Gamma_{ik}({\bf r,r'},\omega)=i\int_0^\infty d\tau e^{i\omega \tau}\langle [E_i(x),E_k(x')]\rangle. \label{4Q}
\end{equation}
The generalized susceptibility is the same as the retarded Green function: $ \Gamma_{ik}(x,x')=G_{ik}^R(x,x')$.  Now take the
 Fourier transform of the two-point function $\langle E_i(x)E_k(x')\rangle$:
 \begin{equation}
  \langle E_i({\bf r}, \omega)E_k({\bf r'},\omega')\rangle= 2\pi \langle E_i({\bf r})E_k({\bf r'})\rangle_\omega \delta (\omega +\omega'), \label{5Q}
  \end{equation}
 where $ \langle E_i({\bf r})E_k({\bf r'})\rangle_\omega$, the  spectral correlation tensor, relates to the retarded Green function as
 \begin{equation}
 \langle E_i({\bf r})E_k({\bf r'})\rangle_\omega={\rm Im} \,G_{ik}^R ({\bf r,r'},\omega)\coth \left( \frac{1}{2}\beta \omega\right),\quad \beta=1/(k_B T) \label{6Q}
 \end{equation}
 (this is the fluctuation-dissipation theorem).
Unless the Green function has an imaginary part, the two-point function is simply {\it zero}. We thus see, also via this alternative route, that absorption is necessary.

Different strategies have been followed in order to deal with the Casimir friction problem. The one followed here will be to adopt a microscopic model, in which we start from the statistical mechanics for harmonically oscillating particles at finite temperature moving nonrelativistically with constant velocity. Such a statistically-mechanical model is of course much more simple than the dielectric model shown in Fig. 1. We argue, however, that such a microscopic model is able to elucidate the essentials of the Casimir friction.

The presentation presented here is based on our three recent papers \cite{hoye10,hoye10a,hoye11}. In turn, they are based on two papers of ours some years ago \cite{hoye92,hoye93}. The microscopic model has been analyzed by other investigators also, in particular by Barton recently \cite{barton10,barton10a,barton11}. We shall have the opportunity to compare with some of his results below.

The other main avenue of approach in order study the friction problem, is to adopt the macroscopic model with permittivity properties. Let us finally refer to some works in that direction: Refs.~\refcite{levitov89,pendry97,pendry10,volokitin08,dedkov08,dedkov10,philbin09}.

\section{Friction force: Statistical mechanical method   }

Consider a quantum mechanical two-oscillator system
whose reference state is the one of uncoupled motion corresponding
to the Hamiltonian $H_0$. The equilibrium situation becomes
perturbed by a time dependent term
 $-Aq(t)$, where $A$ is a time independent
operator and $q(t)$ a classical function of time
whose explicit form depends on the specific properties of the
system. The Hamiltonian becomes $H=H_0-Aq(t)$.
 We put
\begin{equation}
-Aq(t)=\psi ({\bf r}(t))x_1x_2, \label{1}
\end{equation}
where $\psi (\bf r)$ is the  coupling strength (i.e.
$\psi$ is the classical potential between the oscillators). The
separation between the oscillators is   $\bf r$, and $x_1,x_2$ are
the internal vibrational coordinates of the oscillators. When the
oscillators move with respect to each other the coupling has to
vary in time. With nonrelativistic constant relative velocity $\bf
v$ the interaction will vary as
\begin{equation}
-Aq(t)=[ \psi ({\bf r}_0)+{\bf \nabla}\psi({\bf
r}_0)\cdot {\bf v}t+...]x_1x_2, \label{2}
\end{equation}
when expanded around the initial position  ${\bf r=r}_0$ at $t=0$.
 The force between the oscillators, called $\bf B$,  is
\begin{equation}
{\bf B=-(\nabla}\psi ({\bf r}))x_1x_2. \label{3}
\end{equation}
Note that in a mathematical
sense
 the expansion (\ref{2}) requires $vt$ to be small.
Physically, we assume nevertheless Eq.~(\ref{2}) to hold for all
times, so that the interaction energy is  proportional
 to $t$ for all values of $t$. The natural opportunity of choosing
 $q(t)=t$ in the
interaction (\ref{1}) thus has to be modified: as will be shown
below,  a convergence factor will be needed.

Another point worth noticing is that the expression (\ref{1})
corresponds to first quantization only. Quantum electrodynamic
processes such as emission and absorption of photons (second
quantization) are not accounted for by the present model. They
were considered, however,  in Ref.~\refcite{hoye93}.

The equilibrium situation with both oscillators at rest is
represented by the first term in (\ref{2}). It gives rise to a
(reversible) equilibrium force. Thus the friction must be
connected with the second term. To simplify, we will for the
moment neglect the first term. By this the two oscillators will be
fully uncorrelated in their relative  position ${\bf r=r}_0$. The friction force,
 due to the time dependence of the interaction (\ref{2}), will
be a small perturbation upon the equilibrium situation. This
interaction leads to a response $\Delta \langle {\bf B}(t)\rangle$
in the thermal average of $\bf B$. And this is where the Kubo
formula comes in:
\begin{equation}
\Delta \langle {\bf B}(t)\rangle =\int_{-\infty}^t {\boldsymbol
{\phi}}_{BA}(t-t')q(t')dt' \label{4}
\end{equation}
(note that here $\boldsymbol{\phi}_{BA}$ means a vector),
where the response function is given by
\begin{equation}
{\boldsymbol{\phi}}_{BA}(t)=\frac{1}{i\hbar}\rm{Tr} \{ \rho [A,{\bf
B}(t)]\}. \label{5}
\end{equation}
Here $\rho$ is the density matrix and ${\bf B}(t)$ is the
Heisenberg operator ${\bf B}(t)=e^{itH/\hbar}{\bf
B}e^{-itH/\hbar}$, where $\bf B$ like $A$ is time independent. Now
with (\ref{2}) and (\ref{3}), and with $q(t)=t$, expression
(\ref{5}) can be rewritten as $ {\boldsymbol \phi}_{BA}(t)={\bf G}\phi(t),$
with
\[
 {\bf G}=({\bf \nabla}\psi )({\bf v\cdot \nabla}\psi), \quad
 {\bf \phi}(t)=\rm{Tr}\{ \rho \,C(t)\}, \]
\begin{equation}
C(t)=\frac{1}{i\hbar}[x_1x_2,x_1(t)x_2(t)]. \label{7}
\end{equation}
Thus with Eq.~(\ref{4}) and $q(t')=t'$ the force
can be written as
\[ {\bf F}=\Delta \langle {\bf B}(t)\rangle ={\bf
G}\int_{-\infty}^t\phi (t-t')t'dt'={\bf F}_r+{\bf F}_f, \] where
\begin{equation}
{\bf F}_r={\bf G}t\int_0^\infty \phi(u)du \label{8}
\end{equation}
is part of the reversible force by which the part of the force that
represents friction is
\begin{equation}
{\bf F}_f=-{\bf G}\int_0^\infty \phi(u)udu. \label{9}
\end{equation}
Here the new variable $u=t-t'$ has been introduced. The ${\bf
F}_r$ can be interpreted  as a reversible force since it depends only upon position. This interpretation is consistent with the result obtained for the dissipation  below; the ${\bf F}_r$ will not contribute to the net total dissipation.

If one again includes $\psi({\bf r}_0)$ one has
\begin{equation}
 {\bf G}t
=({\bf \nabla}\psi)\psi ({\bf r}_0+ {\bf v}t), \label{10}
\end{equation}
where ${\bf r}={\bf r}_0+{\bf v}t$ is the position at time $t$. By
contrast, expression (\ref{9}) changes sign when the velocity
$\bf v$ changes sign, and it thus represents a friction force.
(Observe that the velocity in (\ref{10}) merely represents the
shift in position.) Equation (\ref{9}) is  the same as
result (2.11) in Ref.~\refcite{hoye92} \footnote{There is a missing minus sign in Eq.~(2.11) and the ${\bf F}_r$ was not taken into account.}, and the Fourier transformed version of it and Eq.~(\ref{9}) above as well, is
\begin{equation}
{\bf F}_f=-i{\bf G}\frac{\partial \tilde{\phi}(\omega)}{\partial
\omega}\Big|_{\omega=0}, \label{11}
\end{equation}
where $\tilde{\phi}(\omega)=\int_0^\infty \phi(t)e^{-i\omega
t}dt~~$ (with $\phi(t)=0$ for $t<0$).

In Ref.~\refcite{hoye92} the Fourier transformed version (\ref{11})
was used to obtain the explicit expression for the friction force.
Here, we instead will use  a different approach based on the  expression (\ref{9}). As in the
reference mentioned we then need the commutator (\ref{7}). This
entity again follows from the properties of quantized harmonic
oscillators. We introduce annihilation and creation operators $a$
and $a^\dagger$ with commutation relations $[a_i, a_i^\dagger]=1$
($i=1,2$; other commutators vanish). As usual, ${a_j} (t)=a_j
e^{-i\omega_jt}$ and ${{a_j}^\dagger(t)}={a_j}^\dagger
e^{i\omega_jt}$. With this the coordinates are
\begin{equation}
x_i=\left(\frac{\hbar}{2m_i\omega_i}\right)^{1/2}(a_i+a_i^\dagger)
\label{12}
\end{equation}
where $m_i$ and $\omega_i$ ($i=1, 2$) are the mass and
eigenfrequency of each oscillator. To obtain $\phi(t)$ from
(\ref{7}) we first have to calculate $ \phi(t)=\langle\langle n_1n_2|C(t)|n_1n_2\rangle\rangle$,
 where $|n_1n_2\rangle =
|n_1\rangle|n_2\rangle$ represents eigenstates with oscillators
excited to levels $n_1$ and $n_2$. Some calculation yields
\begin{equation}
\phi(t)= D \left[(2\langle n_1\rangle +1)\cos
(\omega_1t)\sin (\omega_2t)+(2\langle n_2\rangle+1)\cos
(\omega_2t)\sin (\omega_1t)\right], \label{14}
\end{equation}
where
$ D=\hbar/(2m_1 m_2\omega_1\omega_2) .$
The energy levels are $\varepsilon_n=(n+\frac{1}{2})\hbar \omega$.

In Ref.~\refcite{hoye92} the expression (\ref{14}) was Fourier
transformed to obtain the friction force as given by (\ref{11}).
As an alternative we will here use expression (\ref{14}) directly
in Eq.~(\ref{9}). Then we get the integral \cite{hoye10}
\[ \int_0^\infty te^{-\eta t}\cos(\omega_1t)\sin(\omega_2t)dt \]
\begin{equation}
= \frac{\eta \Omega_1}{(\eta^2+\Omega_1^2)^2}-\frac{\eta
\Omega_2}{(\eta^2+\Omega_2^2)^2} \rightarrow
-\frac{\pi}{2\Omega_2}\delta (\Omega_2), \quad \eta \rightarrow 0.
\label{16}
\end{equation}
Here $\Omega_1=\omega_1+\omega_2$ and
$\Omega_2=\omega_1-\omega_2$. As mentioned above a convergence
factor $e^{-\eta t}$ is needed, and the limit $\eta \rightarrow 0$
is taken. Then the $\Omega_2-$term  becomes a delta function with
prefactor determined by the  integral $
\int_{-\infty}^\infty {\eta
x^2}{(\eta^2+x^2)^{-2}}dx={\pi}/{2}. $
From (\ref{14}) we also get this integral with $\omega_1$ and
$\omega_2$ interchanged. This will  then give the  result (\ref{16})
with opposite sign with respect to the $\Omega_2$-term. Adding up we get the difference of the
prefactors
\[ \coth(\frac{1}{2}\beta \hbar \omega_1)-\coth
(\frac{1}{2}\beta \hbar \omega_2)=-\frac{\sinh (\frac{1}{2}\beta
\hbar \Omega_2)}{\sinh(\frac{1}{2}\beta \hbar \omega_1)\sinh
(\frac{1}{2}\beta \hbar \omega_2)} \]
\begin{equation}
\rightarrow -\frac{\frac{1}{2}\beta \hbar
\Omega_2}{\sinh(\frac{1}{2}\beta \hbar
\omega_1)\sinh(\frac{1}{2}\beta \hbar \omega_2)}, \quad \eta
\rightarrow 0. \label{18}
\end{equation}
Multiplying (\ref{16}) with (\ref{18}) and including the factors $D$
and $\bf G$ the friction force becomes \cite{hoye10}
\begin{equation}
{\bf F}_f=-\frac{\pi \beta \hbar^2({\bf \nabla}\psi)({\bf v\cdot
\nabla}\psi)}{8m_1m_2\omega_1^2\sinh^2(\frac{1}{2}\beta \hbar
\omega_1)}\delta(\omega_1-\omega_2), \label{19}
\end{equation}
which is also the result (3.14) of Ref.~\refcite{hoye92}. Again one
notes that there is friction only when the oscillators have the
same frequency, and $\beta$ should be finite, i.e. $T>0$.

When $\beta \rightarrow \infty$, the expression (\ref{19}) vanishes.
 According to the present oscillator model there is thus no friction force at zero temperature.
 An objection against this result may be that it is somewhat singular due to the presence of the $\delta$-function. Thus its physical significance may not be obvious. However, $\eta$ can be kept finite. This will smooth out the $\delta$-function, and the $\Omega_1$-term in Eq.~(\ref{16}) will give a contribution too. Note that this will not change our conclusions about a finite friction force for $T>0$. (For finite $\eta$, i.e. interaction like a short pulse, there will also be a contribution for $T=0$ due to the $\Omega_1$-term in Eq.~(\ref{16}).) But here we will assume $\eta$ small by which the $T=0$ contribution can be disregarded.

 In Ref.~\refcite{hoye92} the result (\ref{19}) for the friction force
was derived also by two other methods. These methods utilized the
path integral formalism of quantum systems at thermal equilibrium
\cite{hoye81}. The path integral can be identified with a
classical polymer problem where imaginary time is a fourth
dimension of length $\beta$. Thus the polymers stretch out in the
fourth dimension and form closed loops of periodicity $\beta$. For
harmonic oscillators the correlation function along the polymers
is obtained in a straightforward way.
With one of the methods the convolution of
the correlation functions of both oscillators were needed. The
resulting Fourier transform of this convolution was then identified
with the response function $\tilde{\phi}(\omega)$ used in the
expression (\ref{11}) \cite{brevik88}.

With the other method full thermal equilibrium was utilized. Then
the relative motion of the oscillators was regarded as a harmonic
oscillator motion with low frequency $\omega_0\rightarrow 0$.
Again with the path integral one can obtain the Fourier transform
of the response function for the relative motion. The damping of
the relative motion, that can be related to this response
function, gives the friction force, and again the result
(\ref{19}) was recovered. Thus the three methods used in
Ref.~\refcite{hoye92}, as well as the modification considered in the
 present paper, all lead to the same result, in contradiction to some other results in the literature, for instance that of Ref.~\refcite{philbin09}.

The result for the friction force was also extended to the
situation with time-dependent or non-instantaneous interaction
\cite{hoye93}. Then the full thermal equilibrium method was
applicable to generalize the result. With the latter interaction
there was also a friction from the self-interaction of a moving
oscillator with itself.

Consider now the dissipation of energy. It is identified with the work done, during a finite time interval starting from $t=0$ with maximum velocity  $\bf v$ when the
position is ${\bf r=r}_0$. As $t \rightarrow \infty$ the motion is
required to die out. To accomplish this we again introduce the
convergence factor $e^{-\eta t}$ ($\eta \rightarrow 0$), whereby
 $t$ is to be replaced with $q(t)=te^{-\eta t}$.  The velocity decays exponentially,
\begin{equation}
{\bf v} \rightarrow {\bf v}_1(t)={\bf
v}\dot{q}(t)={\bf v}(1-\eta t)e^{-\eta t}. \label{20}
\end{equation}
For $\eta t >0$ , ${\bf v}_1(t)$ will now replace $\bf v$ in expression (\ref{19}) for the friction force. Altogether, the total energy dissipated will be
\begin{equation}
\Delta E=-\int_{-\infty}^\infty {\bf v}_1(t)\cdot
{\bf F}_f\, \dot{q}(t)dt=-{\bf v \cdot F}_f\int_0^\infty
[\dot{q}(t)]^2dt=-\frac{1}{4\eta}{\bf v\cdot F}_f, \label{21}
\end{equation}
where ${\bf F}_f$ is given by Eq.~(\ref{19}).\footnote{The minus signs are missing in Eq.~(21) of Ref.~\refcite{hoye10}.} Note that the
reversible part of the force ${\bf F}_r \propto
t\rightarrow q(t)$ as given by Eq.~(\ref{9}) will not contribute
to the dissipation since $\int_{0}^\infty
\dot{q}(t)q(t)dt=0$.

\section{Energy dissipation calculated from first order perturbation theory}

   Our intention now is to calculate the change $\Delta E$ in energy  by means
   of quantum mechanical perturbation theory; this was considered also in Ref.~\refcite{hoye10a}.  It turns out that the change in energy occurs to second order in the perturbation. Nevertheless, time-dependent perturbation theory to the {\it first} order is sufficient to find this second order effect. This is because the phases of the perturbed change in amplitudes, and the initial amplitudes of the eigenstates, are uncorrelated at thermal equilibrium. Thus change in amplitudes of eigenstates will be the square of perturbed amplitudes; i.e. there are no cross-terms.
   We find that $\Delta E$ is positive,  corresponding to a
   friction force. Doubts occasionally raised in the literature
   about the very existence of the Casimir friction effect
   \cite{philbin09} are thus from this standpoint laid at rest.

    Making use of the  expression for $\Delta E$ we compare the  formalism of the present section with that of Ref.~\refcite{hoye10}, and Sec. 2 above. There,   the linear response via the Kubo formalism was used \cite{hoye92,hoye93} to calculate the force which in turn could be divided into a reversible and an irreversible part.  It is the latter part that is associated with dissipation. A satisfactory feature
   is that the derivation in the present section, although  being  quite different
 from that of Ref.~\refcite{hoye10}, leads to the same physical result.

To fix the notation, we start with perturbation theory for a system at thermal equilibrium. The wave function can be written as
$\psi=\sum_na_n\psi_n, $
where $\psi_n=\psi_n(x)$ are the eigenstates. For simplicity we here let $x$ represent all the coordinates of the system. If $\psi$ is normalized, $\int \psi^*\psi dx=1$, then $|a_n|^2$ is the probability for the system to be in eigenstate $n$. At thermal equilibrium this probability is given by the Boltzmann factor $P_n=|a_n|^2=Z^{-1}\,e^{-\beta E_n}, $
where $E_n$ is the energy eigenvalue of the state and $Z$ is the partition function
$ Z=\sum_ne^{-\beta E_n}.$
Let now the Hamiltonian be perturbed  by the time-dependent interaction
$V(t)=-Aq(t), $
where $A$ is a quantum mechanical operator while $q(t)$ is a scalar function. The $A$ is time independent.

Due to the perturbation the coefficients $a_n$ will change. If the
system starts in a state $m$ there are transitions to other states
given by a change in $a_n$,
$\Delta a_n=b_{nm}.$
The $b_{nm}$ is given by the standard expression
\begin{equation}
b_{nm}=\frac{1}{i\hbar}\int_{-\infty}^tV_{nm}(\tau)e^{i\omega_{nm}\tau} d\tau, \label{6A}
\end{equation}
where \[ V_{nm}(\tau)=\int \psi_n^*\,
V(\tau)\psi_mdx=-A_{nm}\,q(\tau),
\]
$A_{nm}=\langle n|A|m\rangle.$  Here
 $\omega_{nm}=\omega_n-\omega_m$, with $\omega_n=E_n/\hbar$.

As mentioned above, we will assume that the perturbation vanishes after some time. Then we will obtain the total change in $\Delta a_n$ with
\begin{equation}
b_{nm}=-\frac{1}{i\hbar}A_{nm}\,\hat{q}(-\omega_{nm}), \quad
\hat{q}(\omega)=\int_{-\infty}^\infty q(t)e^{-i\omega t}dt,
\label{8A}
\end{equation}
where the hat denotes Fourier transform.

From a general perspective, the system may start in a combination of eigenstates with transitions from several states.  With this,
    $\Delta a_n \rightarrow \sum_{m \neq n}a_mb_{nm}$. Now, the state $n$ does not only receive contributions, but gives away contributions
    to other states also. The latter must follow  from the corresponding increase of probabilities for the other states. Omitting the latter for the moment, the perturbed coefficients
    are
\begin{equation}
a_{1n}=a_n+\Delta a_n=a_n+\sum_{m\neq n}a_mb_{nm}. \label{9A}
\end{equation}
The $a_n$ will have complex phase factors, and in thermal
equilibrium one must assume the phases of $a_n$ and $a_m$ $(m\neq
n$) to be uncorrelated. Thus by thermal average, $\langle a_n^* a_m\rangle =0. $
With this the new probability of the state $n$ becomes
\[
P_{1n}=\langle a_{1n}^* a_{1n}\rangle=|a_n|^2+\sum_{m\neq n}|a_m|^2B_{nm}, \] where $
B_{nm}=b_{nm}b_{nm}^* =|b_{nm}|^2.$
The last term is the increase in probability from the other states. Likewise, the state $n$ must obey a similar loss of probability to other states to conserve probability.
 The loss to other states is thus $\sum_{m\neq n}|a_n|^2B_{mn}$. With Eq.~(\ref{6A}) we have $b_{mn}=b_{nm}^*$, by which $B_{mn}=B_{nm}$. The latter equation reflects that
 the transition probabilities between each pair of states are the
 same in either direction. With this, the resulting perturbed
 probability of state $n$ becomes
 \begin{equation}
  P_{1n}=P_n+\sum_m(P_m-P_n)B_{nm}. \label{12A}
 \end{equation}
 The change in energy can now be evaluated as
 \begin{equation}
  \Delta E=\sum_{nm}E_n(P_m-P_n)B_{nm}
 =\sum_{nm}(E_n-E_m)P_m B_{nm}.
 \label{13A}
 \end{equation}
Utilizing the symmetry with respect to $n$ and $m$ we find
\begin{equation}
\Delta
E=\frac{1}{Z}\sum_{nm}e^{-\frac{1}{2}\beta
(E_n+E_m)}\Delta_{nm}\sinh (\frac{1}{2}\beta \Delta_{nm})B_{nm},
\label{14A}
\end{equation}
with $\Delta_{nm}=E_n-E_m$, and where
\begin{equation}
B_{nm}=(1/{\hbar^2})A_{nm}A_{nm}^*\hat{q}(-\omega_{nm})\hat{q}(\omega_{nm}). \label{C}
\end{equation}

Note that $\Delta E \geq 0$.  The dissipation occurs to
second order in the perturbation. To first order there is no
dissipation; the changes are adiabatic.

\section{Energy dissipation from friction force}

According to the statistical mechanical approach above - Sec. 2 and Ref.~\refcite{hoye10} - the energy dissipation can be written in the form
\begin{equation}
\Delta E=-\int_{-\infty}^\infty v(t)F_f dt=-\int_{-\infty} ^\infty \left[ \int_{-\infty}^t \dot{q}(t)\phi_{AA}(t-t')q(t')dt' \right] dt; \label{21A}
\end{equation}
cf.  Eq.~(27) in Ref.~\refcite{hoye10} (the minus sign in front of the integrals is missing in that reference). The quantity $q(t)$ is most naturally connected with the position, $x(t)=q(t)$, but it can also be interpreted as a
 a position in a more generalized sense as discussed in  Refs.~\refcite{hoye10} and \refcite{hoye10a}, so that the result (\ref{21A}) has a broader applicability. We will now show that this is
  actually the case, by showing that the  result (\ref{21A}) is the
  same as  (\ref{14A}),  obtained by
  means of time-dependent
perturbation theory.

 With wave function representation we first have
 \begin{equation}
 e^{-\beta H} \rightarrow \sum_n\psi_n(x)e^{-\beta
 E_n}\psi_n^*(x_1), \label{22A}
 \end{equation}
 \[
 \rho A A(t)=\frac{1}{Z}\sum_ {nmk}\int \psi_n(x)e^{-\beta
 E_n}\psi_n^*(x_1)A\psi_m(x_1)e^{i\omega_mt}\psi_m^*(x_2)A \]
 \begin{equation}
 \times \psi_k(x_2)e^{-i\omega_kt}\psi_k^*(x_3)dx_1dx_2.
 \label{23A}
 \end{equation}
 Thus we obtain
 \begin{equation}
 {\rm Tr}(\rho AA(t))=\frac{1}{Z}\sum_{nm}e^{-\beta
 E_n}A_{nm}e^{i\omega_mt}A_{mn}e^{-i\omega_nt}, \label{24A}
 \end{equation}
 as $\int \psi_k^*(x)\psi_n(x)dx=\delta _{kn}$ $ (x_3=x_1=x)$, and
 $A_{nm}=\langle n|A|m\rangle$. Likewise we calculate ${\rm
 Tr}(\rho A(t)A)$ by exchange of $\omega_n$ and $\omega_m$ in
 Eq.~(\ref{24A}). The response function becomes
 \begin{equation}
 \phi_{AA}(t)=\frac{1}{i\hbar}{\rm Tr}\left\{ \rho [A, A(t)]\right
 \}=\frac{1}{i\hbar}\sum_{nm}M_{nm}(e^{-i\omega_{nm}t}-e^{i\omega_{nm}t}),
 \label{25A}
 \end{equation}
 with
 \begin{equation}
 M_{nm}=-\frac{1}{Z}e^{-\frac{1}{2}\beta (E_n+E_m)}\sinh (\frac{1}{2}\beta
 \Delta_{nm})A_{nm}A_{nm}^* \label{26A}
 \end{equation}
 (recall that $\Delta_{nm}=E_n-E_m=\hbar \omega_{nm},
 ~A_{mn}=A_{nm}^*)$.  The expression for $M_{nm}$ follows
 if  one first exchanges  $n$ and $m$ in Eq.~(\ref{24A}), then adds the
 resulting term to it and divides by 2. Some manipulation then yields
 \begin{equation}
 \Delta E=\frac{1}{\hbar}\sum_{nm}M_{nm}\,\omega \,
 \hat{q}\,(\omega)\,\hat{q}(-\omega). \label{29A}
 \end{equation}
 With $\omega=\omega_{nm}=\Delta_{nm}/\hbar$ and $M_{nm}$ given by
 the expression (\ref{26A}) this is nothing but the result
 (\ref{14A}), but now  obtained by time-dependent
 perturbation theory. Thus we have been able to derive the same expression for the dissipated energy in two independent ways.

   The result (\ref{14A})  (or (\ref{29A})) may be applied to the pair of interacting harmonic oscillators considered in Sec. 2, but we abstain  from further details here; the reader is referred to Ref.~\refcite{hoye10a}. We give, however, the final expression for
$\Delta E$:
\begin{equation}
\Delta E=\frac{\pi \beta \hbar^2\gamma^2}{8\eta \sinh^2(\frac{1}{2}\beta \omega_1)}\delta (\omega_1-\omega_2), \label{A}
\end{equation}
where
\begin{equation}
\gamma=\left(\frac{1}{2}D\hbar \right)^{1/2}({\bf v \cdot \nabla}\psi), \label{B}
\end{equation}
with $D$ given below Eq.~(\ref{14}). This is the same as the result (\ref{21}), with expression (\ref{19}) inserted.

\bigskip

We may summarize our developments of this section:

1.  The total energy dissipation was calculated for a system perturbed by a varying interaction. The change in energy is basically a second order effect but was calculated with the use of standard time-dependent perturbation theory to first order only, the reason being the absence of cross terms due to uncorrelated phases of eigenstates. The energy change was found to be positive or zero. The result agrees with our previous results of Ref.~\refcite{hoye10}, obtained in a different and independent way.

2. We have in general assumed finite temperature, and initial thermal equilibrium. Moreover, we have assumed low velocities and nonrelativistic mechanics, whereby photons are not present. Photons were introduced, however, in our earlier study \cite{hoye93}.

3.  The energy change $\Delta E$ is finite in general. This corresponds to a finite friction force. In the limit $T\rightarrow 0$ our formalism gives, however, that  $\Delta E\rightarrow 0$ for the model considered in Sec. 2. This result is due to our assumption about constant velocity, involving slowly varying coupling. For couplings varying more rapidly, there will also be a friction force at $T=0$, due to transitions to excited states.

\section{ Equivalence between different formulations}

As mentioned, there are different formulations in the literature on how to deal with Casimir friction. In the present section, assuming $T=0$,  we wish to compare our formalism above with that recently given by Barton \cite{barton10,barton10a} (cf. also Ref.~\refcite{barton11} dealing with finite temperature). In particular, Ref.~\refcite{barton10}
 dealt with a two-oscillator model, thus in essence the same microscopic model as ours. Barton analyzed the system using quantum mechanical perturbation theory. The striking point is that the expressions he obtained are seemingly quite different from those we obtained in Ref.~\refcite{hoye10}. In Barton's own words (Ref.~\refcite{barton10}, Sect. 3) "..in view of the manifold current controversies about quantum-governed frictional force generally, it seems well worth exploring whether such differences reflect substantiate disagreement or only a confusion of terms".
Our present investigation is  a follow-up of Barton's suggestion. We intend to demonstrate explicitly that the obtained expressions for the dissipated Casimir energy are in fact  in agreement with each other, thus a reassuring result.

 Assume that the oscillators have the same eigenfrequency $\omega$ and the same mass $m$. They interact via a time-dependent Hamiltonian
\begin{equation}
H_{\rm int}=\frac{e^2}{s^3}\,y_1y_2, \label{1C}
\end{equation}
(Gaussian units assumed).  Here $e$ is the elementary charge, $y_1$ and $y_2$ the oscillator coordinates, and ${\bf s =s}(t)$ is the  vectorial distance between the mass centers, varying with time because of the relative motion of the oscillators.  Introducing new coordinates
\begin{equation}
y_\pm =\frac{y_1 \pm y_2}{\sqrt 2}, \label{2C}
\end{equation}
one can write the interaction Hamiltonian as
\[ H_{\rm int\pm}=H_{\rm int+}+H_{\rm int-}, \]
\begin{equation}
H_{\rm int\pm}=\pm \frac{1}{2}q\,y_\pm^2, \quad q=\frac{e^2}{s^3}. \label{3C}
\end{equation}
By use of time-dependent perturbation theory the total energy dissipated is then found to be
\[ \Delta E=2\times 2\hbar \omega |c(\infty)|^2,  \]
\begin{equation}
c(t)=-\frac{i}{2\hbar}\int_{-\infty}^t dt' q \,\langle 2_+ |y_+^2|0_+\rangle \, e^{2i\omega t'}, \label{4C}
\end{equation}
as given by Eqs.~(3.3) and (2.4) respectively, in Ref.~\refcite{barton10}. At $T=0$,  only excitations from the ground state are possible.

To compare Barton's result (\ref{4C}) with ours, we first have to evaluate the matrix elements in (\ref{4C}). In terms of the creation and annihilation operators we have ($\omega_\pm \rightarrow \omega$ for small perturbations)
\begin{equation}
y_\pm= \sqrt{b}\,(a_\pm+a_\pm^\dagger),\quad b=\frac{\hbar}{2m\omega}. \label{5C}
\end{equation}
Then,
\begin{equation}
\langle 2_+ |y_+^2 |0_+ \rangle=b\,\langle 2_+ |{a_+^\dagger}^2 |0_+ \rangle =\sqrt{2} \,b. \label{6C}
\end{equation}
Together with Eq.~(\ref{4C}) this gives
\[ \Delta E=8\hbar \omega b^2 |I(\infty)|^2, \]
\begin{equation}
I(t)=-\frac{i}{2\hbar}\int_{-\infty}^t dt' q\, e^{2i\omega t'}. \label{7C}
\end{equation}
Proceed now to compare this result with those that we derived in Refs.~\refcite{hoye10} and \refcite{hoye10a}, at $T=0$.  The interaction Hamiltonian,  written as $H_{\rm int}=-Aq(t)$, now corresponds to
\begin{equation}
A=-y_1y_2, \quad  {\rm and} \quad q(t)=q=\frac{e^2}{s^3}. \label{8C}
\end{equation}
Recall that  $A$ is a time-independent operator accounting for the quantum mechanical properties of the system, while $q(t)$ is a classical function of time. At $T=0$ the system is in its ground state with probability equal to one.

The change in energy is given by  Eq.~(\ref{13A}),  where $E_n$ is the energy in the (unperturbed) eigenstate $n$, and $P_n=|a_n|^2$ is the probability of the system to be in this state. As mentioned, we start from the ground state so that $P_m \rightarrow P_{00}=1$. Further, as $B_{nm}=|b_{nm}|^2$ with $b_{nm}$ the transition coefficient between states $m$ and $n$,  Eq.~(\ref{13A}) reduces to
\begin{equation}
\Delta E=(E_{11}-E_{00})B_{1100}, \label{10C}
\end{equation}
where $E_{11}-E_{00}=2\hbar \omega$ is the energy difference between the state $|11\rangle$ where both oscillators are excited to the first level, and the ground state $|00 \rangle$. The coefficient $B_{1100}$ is the transition probability between these two states.

What remains is to calculate $B_{1100}$. Using  Eq.~(\ref{8A}) we get
\begin{equation}
\hat{q}(-\omega_{nm})\rightarrow \hat{q}(-\omega_{1100})=\hat{q}(-2\omega)=2i\hbar I(\infty), \label{12C}
\end{equation}
with $I(\infty)$ given by Eq.~(\ref{7}). Further,
\[ A_{nm} \rightarrow A_{1100} =\langle 11|-y_1y_2|00\rangle \]
\begin{equation}
=-\langle 1|y_1|0\rangle \langle 1|y_2|0\rangle=-\langle 1|y_1|0\rangle^2=-b\langle 1|a_1^\dagger |0\rangle^2 =-b. \label{13C}
\end{equation}
Altogether, when inserted into Eq.~(\ref{10C}) we obtain
\[ B_{1100}=\frac{1}{\hbar^2}|A_{1100}|^2\hat{q}(-2\omega)\hat{q}(2\omega)=4b^2|I(\infty)|^2, \]
\begin{equation}
\Delta E=8\hbar \omega b^2|I(\infty)|^2, \label{14C}
\end{equation}
which coincides with the result obtained by Barton, Eq.~(\ref{4C}) above.

As an additional remark we note that the situation with zero friction at $T=0$ for slowly varying forces can be analyzed in a straightforward way from the equations above. With time-dependent part ($q=0$, $t<0$),
\begin{equation}
q(t)= te^{-\eta t}, \quad t>0,
\label{15C}
\end{equation}
 we namely find when $\eta \rightarrow 0$
\begin{eqnarray}
\hat q(\omega)&=&\frac{1}{(\eta+i\omega)^2},\\
\hat q(\omega)\hat q(-\omega)&=&\frac{1}{(\eta^2+\omega^2)^2}\rightarrow \frac{\pi}{2\eta\omega^2} \delta(\omega).
\label{16C}
\end{eqnarray}
 Thus with $\omega\neq 0$, Eq.~(\ref{16C})  will give zero for the dissipated energy $\Delta E$.

 [By looking at this in some more detail it would seem that the middle term in Eq.~(\ref{16C}), before taking the limit
  $\eta\rightarrow 0$, implies there to be some  dissipation. This is physically an artefact, due to our assumption of an abrupt change of  $q(t)$ at $t=0$.]

The equivalence between Barton's results and ours is therewith shown. We thus hope to have shed light on one of the subtle issues in the  Casimir friction world demonstrating that quite different approaches can lead to the same result.

\section*{Acknowledgment}

I.B. thanks Gabriel Barton for valuable correspondence.



\end{document}